\documentclass[aps,prb,twocolumn,superscriptaddress,showpacs]{revtex4}
\usepackage{amsmath}
\usepackage{graphicx}

\begin{document}

\title{Physical properties of ferromagnetic-superconducting coexistent system}

\author{Hari P. Dahal}
\affiliation{Department of Physics, Boston College, Chestnut Hill, MA, 02467}

\author{Jason Jackiewicz}
\affiliation{Department of Physics, Boston College, Chestnut Hill,
MA, 02467}

\author{Kevin S. Bedell}
\affiliation{Department of Physics, Boston College, Chestnut Hill, MA, 02467}

\date{\today}

\begin{abstract}
We study the nuclear relaxation rate $1/T_{1}$ of a
ferromagnetic-superconducting system from the mean field model
proposed in Ref.$14$. This model predicts the existence of a set
of gapless excitations in the energy spectrum which will affect
the properties studied here, such as the density of states and,
hence, $1/T_{1}$.  The study of the temperature variation of
$1/T_{1}$ (for $T<T_{c}$) shows that the usual Hebel-Slichter peak
exists, but will be reduced because of the dominant role of the
gapless fermions and the background magnetic behavior. We have
also presented the temperature dependence of ultrasonic
attenuation and the frequency dependence of electromagnetic
absorption within this model. We are successful in explaining
certain experimental results.
\end{abstract}

\pacs{71.27.+a, 75.10.LP, 76.60.Jx, 76.60.-k}

\maketitle

\section{\label{sect-model} Introduction}

In the BCS theory of superconductivity\cite{bcs1957},the
conduction electrons in a metal cannot be both ferromagnetically
ordered and superconducting. Superconductors expel magnetic field
passing through them but strong magnetic fields kill the
superconductivity(SC). Even small amounts of magnetic impurities
are usually enough to eliminate SC. Much work has been done both
theoretically and experimentally to understand this interplay and
to search for the possibility of coexistence between these two
ordered states.

In the conventional theory of SC, ferromagnetism(FM) and SC
compete with each other but in principle it is possible for any
metal to become a SC in its non-magnetic state at a sufficiently
low temperature. Even strongly ferromagnetic iron
\cite{katsuya2001} undergoes a superconducting transition at low
temperature under application of sufficient pressure to bring it
to its nonmagnetic state. But interest is in simultaneous
existence of both of the ordered states. A. Abrikosov
\cite{abrikosov1988} studied superconductivity with magnetic
impurities using the RKKY interaction in which magnetic impurities
interact with conduction electrons with the magnetization
considered as an external parameter independent of the
superconducting gap, and showed that the normal ferromagnetic
state has lower energy than the SC-FM state and hence coexistence
is energetically unfavorable. Fulde and Ferrell \cite{fulde1964}
studied superconductivity with a strong exchange field produced by
ferromagnetically aligned impurities and found that if the
exchange field is sufficiently strong compared to the energy gap,
a new type of depaired superconducting ground state will occur.
Fay and Appel \cite{fay1980} predicted the possibility of a p-wave
superconducting state in itinerant ferromagnets where the pairing
is mediated by the exchange of longitudinal spin fluctuations.
They showed that when the ferromagnetic transition is approached
from either the ferromagnetic or paramagnetic side, the p-wave
transition temperature goes through a maximum and then falls to
zero. Even if superconductivity is interpreted as arising from
magnetic mediation, it was thought that the SC state will occur in
the paramagnetic phase. But magnetically mediated
superconductivity was never observed \cite{coleman2000}. Some
theories predicted the possibility of s-wave SC and FM coexistent
order in the paramagnetic phase but the ferromagnetic fluctuation
destroys it near the Curie temperature. Consideration of s-wave
superconductivity and ferromagnetism has been carried out by
Suhl\cite{suhl2001} and Abrikosov\cite{abrikosov2001} in which the
ferromagnetism is due to localized spins. There had been no
experimental observations of coexistence until it was found in the
ferromagnetic metal $UGe_{2}$\cite{saxena2000}. The coexistence
has also been shown to exist in $ZrZn_{2}$ \cite{pfleiderer2001}
and $URhGe$ \cite{aoki2001}. Experiments on these three materials
show that the same electrons are responsible for both ordered
states. But still the exact symmetry of the paired state and the
dominant mechanism responsible for the pairing is in question.
Although most authors believe there is triplet superconductivity
in these materials, the possibility of s-wave superconductivity
cannot be denied.

Blagoev \textit{et al.}\cite{blagoev1998,blagoev1999} studied a
weak ferromagnetic Fermi-liquid and showed that s-wave
superconductivity is possible and favored on the ferromagnetic
side. With similar thought, Karchev \textit{et
al.}\cite{karchev2001} developed an itinerant ferromagnetic model
in a mean field approach in which the magnetic electrons are also
the one responsible for the formation of the Cooper pairs.
Following this model, two of the authors \cite{jackiewicz} of this
paper studied the specific heat and compared it with the
experimental data of $UGe_{2}$. They have shown that this model
exhibits the quantitative behavior of the specific heat. The phase
diagram shown by them is similar to that found in $UGe_{2}$
experimentally. So there is enough room to believe the existence
of s-wave superconductivity in the coexistence state. Motivated
with these thoughts, in this paper we are going to present the
results of calculations of the density of states, nuclear
relaxation rate, ultrasonic attenuation and electromagnetic
absorption in the model given by ref.\cite{karchev2001}.

We refer the reader to ref.\cite{karchev2001} for the details, and here
we look at the mean-field
Hamiltonian obtained from a model Hamiltonian by the standard mean-field procedure,

\begin{eqnarray}
\label{mf-ham}
\nonumber
 H_{mf}&=&\sum_{\vec{p}}\epsilon_{p}
(c_{\vec{p}}{}_{\uparrow}^{\dag}
c_{\vec{p}}{}_{\uparrow}+ c_{\vec{p}}{}_{\downarrow}^{\dag}c_{\vec{p}}{}_{\downarrow})\\
\nonumber
      &+&\frac{JM}{2}\sum_{\vec{p}}(c_{\vec{p}}{}_{\uparrow}^{\dag}
c_{\vec{p}}{}_{\uparrow}- c_{\vec{p}}{}_{\downarrow}^{\dag}c_{\vec{p}}{}_{\downarrow})\\
      &-&\sum_{\vec{p}}(\Delta c_{\vec{p}}{}_{\uparrow}^{\dag}c_{-\vec{p}}
{}_{\downarrow}^{\dag}+H.c.)+\frac{1}{2}JM^{2}+\frac{|\Delta|^{2}}{g}.
\end{eqnarray}
\noindent

The diagonalization of this Hamiltonian using a Bogoliubov
transformation yields,

\begin{equation}
H_{MF}=E_{0} + \sum_ {\vec{p}}\left(E_{p}^{\alpha}\alpha _{\vec{p}}^{\dagger}
\alpha _{\vec{p}}+E_{p}^{\beta}\beta _{\vec{p}}^{\dagger}
\beta _{\vec{p}}\right),
\end{equation}
\noindent where
\begin{eqnarray}
\nonumber E_{0} =
\sum_{\vec{p}}\epsilon_{\vec{p}}^{\downarrow\uparrow} +
\frac{1}{2}JM^{2}+\frac{|\Delta|^{2}}{g},  \\
\nonumber \epsilon_{\vec{p}}^{\downarrow}{}^{\uparrow} =
\frac{p^{2}}{2m^{*}} - \mu \mp \frac{JM}{2}.
\end{eqnarray}

\noindent The quasiparticle energy dispersion relations are,
\begin{eqnarray}
\label{dispersion}
E_{p}^{\alpha} = \frac{JM}{2} + \sqrt{\xi_{p}^{2} + |\Delta|^{2}}, \\
E_{p}^{\beta}  = \frac{JM}{2} - \sqrt{\xi_{p}^{2} + |\Delta|^{2}}.
\end{eqnarray}
\noindent

The final step is to minimize the free energy to produce the
mean-field equations.  This results in a set of two coupled
equations in $M$ and $\Delta$ that will be solved
self-consistently below. For $M$ we find,

\begin{equation}
\label{total-mag}
M=\frac{1}{2}\int\frac{d^{3}p}{(2\pi)^3}(1-n_{p}^{\alpha}-n_{p}^{\beta}),
\end{equation}
\noindent and for $\Delta$,
\begin{equation}
\label{total-gap}
|\Delta|=\frac{|\Delta| g}{2}\int\frac{d^{3}p}{(2\pi)^3}\frac{n_{p}^{\beta}
-n_{p}^{\alpha}}{\sqrt{\xi_{p}^{2} + |\Delta|^{2}}}.
\end{equation}

The two order parameters, $M$ and $\Delta$, have dependencies such
as $M=M(g,J,T)$ and $\Delta=\Delta(g,J,T)$. Moreover they are
coupled with each other through the distribution functions which
are also functions of $M$ and $\Delta$.  From the numerical
solutions of these two coupled equations
(\ref{total-mag},\ref{total-gap}),we will get the values of M and
$\Delta$ to be used for the calculation of physical parameters we
are interested in. We want to clarify that all of the future
results are derived strictly from the dispersion relations and the
mean-field equations only, with no other assumptions made about
the coupling strength limits or small magnetization. In what
follows we have studied only the case when $JM>2\Delta$ by which
we mean that there already exists weak ferromagnetic order in
which arises superconductivity.

\section{\label{sect-theoryone} ENERGY SPECTRUM AND DENSITY OF STATES}

If we compare the quasi-particle energy spectra Eqs.(3,4) with the
BCS energy spectrum $E=\sqrt{\xi_{p}^{2} + |\Delta|^{2}}$, we see
that the modification is due to the presence of the ferromagnetic
energy. The plot of the energy spectrum as a function of $\xi_{p}
= \frac{{p}^{2}}{2{m}^{\ast}}-\mu $, the energy of an excited
particle above the Fermi level, is shown in Fig.1. It is very
clear that the gap is not at the Fermi level, rather it is pushed
up. Now the energy spectrum is symmetric around $E=\frac{JM}{2}$,
contrary to the BCS case where it was symmetric around $E=0$. The
fermions which follow the different energy dispersions have
different properties. The fermions which follow the
$E_{p}^{\alpha}$ energy spectrum are BCS-like excitations, and
will be called alpha fermions hereafter,and have a gap in the
energy. Those fermions which follow the $E_{p}^{\beta}$ energy
spectrum do not have a gap and will be called beta fermions or the
gapless excitations hereafter. The presence of these gapless
fermions will change the physical properties of the system. The
maximum of the energy of beta fermions is $JM/2 -\Delta$ and the
minimum of the energy of the alpha fermions is $JM/2 + \Delta$ and
so there will be a gap of $2\Delta$. But at $T=0$, the beta
fermions fill only up to $E=0$, so the gap of alpha fermions is
$JM/2 + \Delta$.

 \begin {figure}
\includegraphics[width=9.2 cm,height=6.8cm]{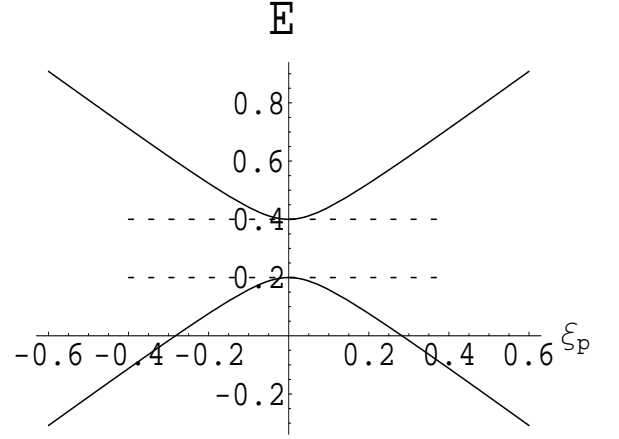}
\caption{Energy spectrum of excited fermions. The upper curve
corresponds to alpha fermions and lower one is for beta fermions.
The upper horizontal line is at $JM/2+\Delta$ and the lower
horizontal line is at $JM/2-\Delta$. The gap is $2\Delta$. Beta
fermion excitations are also associated with the zero energy
excitations called gapless fermions.}
\end{figure}

Next, we derived the expressions for the density of states for the
corresponding fermions. We use the usual relation,
\begin{equation}
\begin{split}
N^{i}(E)= \frac{1}{{2\pi}^{3}}\int{d\bar{p} \delta(E-E_{p}^{i})}.
\end{split}
\end{equation}
where $i$ refers to alpha and beta fermions. Using the property of
the delta function and solving the equation, we get,
\begin{equation}
\begin{split}
\frac{N^{\beta}(E)}{N(0)}=
\frac{1}{4\pi^{2}}[\frac{{\sqrt{{p_{F}^{2}+2m^{*}\sqrt{(\frac{JM}{2}-E)^{2}-\Delta^{2}}}}
 }}{|[\frac{1}{2m^{*}}\frac{\sqrt{(\frac{JM}{2}-E)^{2}-\Delta^{2}}}{\frac{JM}{2}-E}]|}\\
 +\frac{{\sqrt{{p_{F}^{2}+2m^{*}\sqrt{(\frac{JM}{2}-E)^{2}-\Delta^{2}}}}}}
 {|[-\frac{1}{2m^{*}}\frac{\sqrt{(\frac{JM}{2}-E)^{2}-\Delta^{2}}}{\frac{JM}{2}-E}]|}].
 \label{denbeta}
 \end{split}
\end{equation}

\begin{equation}
\begin{split}
\frac{N^{\alpha}(E)}{N(0)}=
\frac{1}{4\pi^{2}}[\frac{{\sqrt{{p_{F}^{2}+2m^{*}\sqrt{(E-\frac{JM}{2})^{2}-\Delta^{2}}}}
 }}{|[-\frac{1}{2m^{*}}\frac{\sqrt{(E-\frac{JM}{2})^{2}-\Delta^{2}}}{E-\frac{JM}{2}}]|}\\
 +\frac{{\sqrt{{p_{F}^{2}+2m^{*}\sqrt{(E-\frac{JM}{2})^{2}-\Delta^{2}}}}}}
 {|[\frac{1}{2m^{*}}\frac{\sqrt{(E-\frac{JM}{2})^{2}-\Delta^{2}}}{E-\frac{JM}{2}}]|}].
 \label{denalpha}
 \end{split}
\end{equation}
Both of the expressions converge to the density of normal fermions
at the Fermi level, $N(0) =\frac{mp_{f}}{\pi^{2}}$, in the
corresponding limit. And, for $M=0$, the expressions converge to
to the density of states of the BCS model, $N(E)/N(0) =
E/\sqrt{(E^{2} - \Delta^{2})}$. Both of the expressions are the
same mathematically, the only difference lies in the fact that the
energy for gapless fermions ranges from $-\infty$ to
$JM/2-\Delta$, including zero obviously and that for alpha
fermions ranges from $JM/2+\Delta$ to $+\infty$. The result of the
calculation of normalized density of states with respect to
density of states of normal state fermions at the Fermi level is
as shown in Fig.2.

\begin{figure}
\includegraphics[width=8.0 cm,height=6.8cm]{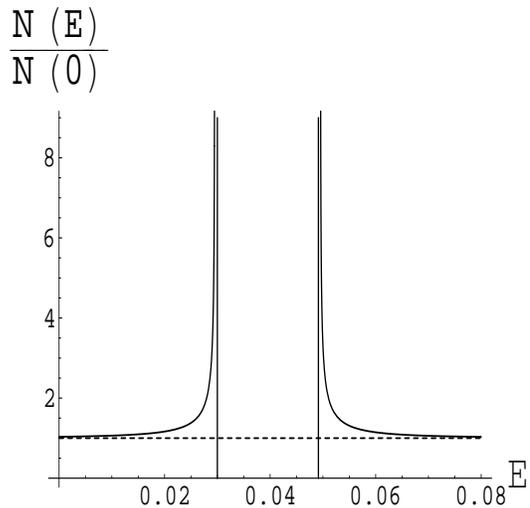}
 \caption{Density of states as a function of Energy of excited fermions in our model.
  There is finite density of gapless fermions. The density of
  state is diverging at $JM/2+\Delta$ and $JM/2-\Delta$.
  The gap is  $2\Delta$.}
\end{figure}

As we have mentioned above, both expressions for the density of
states are identical. If we plot both the expression from
$-\infty$ to $+\infty$, the plot will be same. So we can just use
any one of the density of states expressions wherever needed,
preserving the limit of energy ranges suitable for the
corresponding fermions. In the above plot, the left most curve
corresponds to the beta fermion densities and the right most is
for the alpha fermions. The gap is pushed up (compared to the BCS
gap) entirely in the positive energy side giving rise to a finite
density for the gapless fermions at the original Fermi level. The
density of states is enhanced at $JM/2-\Delta$ for beta fermions
and at $JM/2+\Delta$ for alpha fermions and has no density in the
gap of $2\Delta$. Looking at the energy spectrum and the density
of states, we can guess that the physical properties of the system
will be dominated by normal metal like behavior due to the
presence of gapless fermions at the Fermi energy.

In order to get a good feeling about the type of gapless fermions
we are dealing with, it is better to compare this with gapless
fermions discussed in the context of a superconductor with a
magnetic impurity. The origin of gapless fermions in FS (our
model) and a magnetic impurity superconductor is different. In
magnetic impurity superconductors for zero impurity concentration,
there will be full fermions condensation, full gap and diverging
density of states at the edge of gap energy. But an increase in
impurity concentration brings de-pairing effects in Cooper pairs.
The interaction of fermions with the impurity triggers a flip of
the spin of the Cooper pair giving rise to what is called a pair
breaking effect. The fermions resulting from the pair breaking
mechanism will start to fill in the gap. The diverging density of
states starts to smear out resulting in a finite density of states
for all the energy. At certain impurity density the gap will be
fully populated giving rise to gapless fermions. The zero energy
gap does not necessarily mean that there is no superconductivity
because the order parameter will not be zero together with the gap
for that impurity concentration. But the transition temperature
will be reduced indicating a suppressed superconductivity. The
superconductivity will be completely suppressed at a certain
critical impurity density which is bigger than the density at
which the gap will be zero\cite{abrikosov1988}. So the gaplessness
is due to the de-pairing effect of the Cooper pair. But in our
model of FS we see that the presence of gapless fermions is the
inherent property of FS. Neither the gap should be zero nor should
the density of states be smeared out for the gapless fermions to
exist. There is always Fermi surface where gapless excitations are
present no matter what the value of the gap is. Their existence is
independent of each other.

Even though the origin of gapless fermions is different in these
two different cases, their presence has a common effect in
reducing the transition temperature. In our model of FS the
presence of gapless fermions is necessary for the magnetic
property of the system and hence will reduce the available
fermions to pair up. This reduces the transition temperature.
Mathematically we can understand how the transition temperature is
reduced due to the presence of magnetism. Let us compare the
equation for the gap parameter in the BCS and FS cases. For the
BCS case,
\begin{equation}
\frac{1}{gN(0)}=
\int_{0}^{\lambda}{\frac{d\epsilon}{\sqrt{\epsilon^{2}+|\triangle|^{2}}}}
\end{equation}
where $\lambda$ is the cutoff value. In the weak coupling limit,
the equation can be solved to get an expression for the gap which
looks like,
\begin{equation}
\triangle= 2{\lambda}{\exp^{(\frac{-1}{gN(0)})}}
\end{equation}
For our model of FS, referring to Eq.(14)\cite{karchev2001}, the
equation (10) above looks like,
\begin{equation}
\frac{1}{gN(0)}=
\int_{\sqrt{\frac{JM}{2}^{2}-\Delta^{2}}}^{\lambda}{\frac{d\epsilon}{\sqrt{\epsilon^{2}+|\triangle|^{2}}}}
\end{equation}
and in the weak coupling limit, the gap equation reads as,
\begin{equation} \triangle=
2{\lambda}{\exp^{-(\frac{1}{gN(0)}+{\sqrt{(\frac{JM}{2})^{2}-
\Delta^{2}}})}}
\end{equation}
Analyzing these expressions we clearly see that the presence of
the magnetization reduces the volume of phase space that is
available for the Cooper pair. This leads to a decrease in the gap
at $T=0$ and hence $T_{c}$ will also be reduced compared to the
BCS value for given values of other relevant parameters. So the
increase in magnetization pushes the gap away from the Fermi
surface and reduces its magnitude also.

Next we study some interesting properties of the coexistent system
such as nuclear relaxation, ultrasonic attenuation and
electromagnetic absorption following the calculation done by many
authors for BCS
model.\cite{tinkham1996,hslichter1975,hebelslichter1959,hebel1959}.

\section{\label{sect-theorytwo} TRANSITION PROBABILITIES}

To study transition probabilities we need to find out the
expressions for the coherence factors appropriate for different
cases. For that we use the coefficients of the Bogoliubov
transformations, namely $u$ and $v$,

\begin{equation}
\begin{split}
|v_{k}|^{2}=1-|u_{k}|^{2}=\frac{1}{2}(1-\frac{\xi}{\sqrt{\xi^{2}+|\Delta|^{2}}})
\end{split}
\end{equation}

Then the expressions for the coherence factors for the beta
fermions and alpha fermions are,

\begin{equation}
F^{\beta,\alpha}=(uu^{\prime}\mp vv^{\prime})^{2}=
\frac{1}{2}(1\mp\frac{|\Delta|^{2}}{(\sqrt{\xi^{2}+
|\Delta|^{2}}){(\sqrt{\xi^{2}+|\Delta|^{2}})}^{\prime}})
\end{equation}
for the scattering process, and

\begin{equation}
F^{\beta,\alpha}=(vu^{\prime}\pm uv^{\prime})^{2}=
\frac{1}{2}(1\pm\frac{|\Delta|^{2}}{(\sqrt{\xi^{2}+
|\Delta|^{2}}){(\sqrt{\xi^{2}+|\Delta|^{2}})}^{\prime}})
\end{equation}
for the creation or annihilation process. If we express in terms
of quasi-particles energy explicitly, these equations reduce to

\begin{equation}
\begin{split}
F^{\beta\beta}(\Delta,E,E^{\prime})=\frac{1}{2}(1\mp\frac{|\Delta|^{2}}{(\frac{JM}{2}-E)(\frac{JM}{2}-E^{\prime})})
\end{split}
\end{equation}

\begin{equation}
\begin{split}
F^{\alpha\alpha}(\Delta,E,E^{\prime})=\frac{1}{2}(1\mp\frac{|\Delta|^{2}}{(E-\frac{JM}{2})(E^{\prime}-\frac{JM}{2})})
\end{split}
\end{equation}

\begin{equation}
\begin{split}
F^{\beta\alpha}(\Delta,E,E^{\prime})=\frac{1}{2}(1\mp\frac{|\Delta|^{2}}{(\frac{JM}{2}-E)(E^{\prime}-\frac{JM}{2})})
\end{split}
\end{equation}
for the scattering process, and

\begin{equation}
\begin{split}
F^{\beta\beta}(\Delta,E,E^{\prime})=\frac{1}{2}(1\mp\frac{|\Delta|^{2}}{(\frac{JM}{2}-E)(\frac{JM}{2}-E^{\prime})})
\end{split}
\end{equation}

\begin{equation}
\begin{split}
F^{\alpha\alpha}(\Delta,E,E^{\prime})=\frac{1}{2}(1\pm\frac{|\Delta|^{2}}{(E-\frac{JM}{2})(E^{\prime}-\frac{JM}{2})})
\end{split}
\end{equation}

\begin{equation}
F^{\beta\alpha}(\Delta,E,E^{\prime})=\frac{1}{2}(1\pm\frac{|\Delta|^{2}}{(\frac{JM}{2}-E)(E^{\prime}-\frac{JM}{2})})
\end{equation}
for the pair creation or annihilation process. It is clear now
that $F$ is like a matrix with diagonal components for intra band
transitions and off diagonal components for inter band
transitions. The scattering process which prefers the upper sign
is called case I and the process which prefers the lower sign is
called case II. In the low frequency limit these corresponds to
ultrasonic attenuation and nuclear relaxation respectively.

The effect of the coherence factors will be discussed with respect
to the properties studied here. We are now ready to have an
expression for the net transition rate $\frac{1}{T_{1}}$, for the
coexistent state, between energy levels E and $E'= E +
\hbar\omega$, and is expressed below as,

\begin{equation}
\frac{1}{T_{1}} = |M|^{2}\int{F(\Delta,E,E')N_{s}(E)N_{s}(E')
(f(E)-f(E'))dE}
 \end{equation}
where $M$ is the magnitude of a one-electron matrix element. Since
we are interested in the ratio to the normal state scattering
rate, we do not need to know more about the actual value of $M$.
The limit of integration is from $-\infty$ to $+\infty$ with an
obvious understanding that the limits of the integral have a gap
of $2\Delta$, from $JM/2 - \Delta$ to $JM/2 + \Delta$, if we are
working with finite temperature but the maximum limit of the
integral will be only zero if we are working at zero temperature
since $f(E)$ and $f(E^{\prime})$ both will be zero for $E>0$. The
coherence factors will be chosen appropriate to the band involved
in the transition. $f(E)$ is the usual probability distribution
function. We will study the ratio of this transition rate with
respect to that of the normal state fermions at the Fermi level
$(\frac{1}{T_{n}}=|M|^{2}N^{2}(0)\hbar\omega)$.

In the limit of $\hbar\omega\rightarrow 0$ there will be
transitions due only to scattering and we do not need to consider
inter-band transitions. This applies to nuclear relaxation and
ultrasonic attenuation but for electromagnetic absorption we can
vary the frequency to higher values so that the inter-band
transition will also contribute to the scattering.

\section{\label{sect-theorythree} ULTRASONIC ATTENUATION}

The relevant matrix elements for treating the attenuation of
longitudinal sound waves have case I scattering coherence factors.
So in the limit of $\hbar\omega\rightarrow 0$ the ratio of the
transition rate $(T_{n}/T_{1})$ can be expressed as,

\begin{equation}
\begin{split}
\frac{T_{n}}{T_{1}}
=\int_{-\infty}^{\frac{JM}{2}-\Delta}{(1-\frac{\Delta^{2}}{({\frac{JM}{2}}-E)^{2}})}
{(\frac{N^{\beta}(E)}{N(0)})^{2}}{(\frac{-\partial{f(E)}}{\partial{E}})}dE\\
 +\int_{\frac{JM}{2}+\Delta}^{\infty}{(1-\frac{\Delta^{2}}{(E-\frac{JM}{2})^{2}})}
{(\frac{N^{\alpha}(E)}{N(0)})^{2}}{(\frac{-\partial{f(E)}}{\partial{E}})}dE.
 \label{ultrasonic}
 \end{split}
\end{equation}

We did the numerical calculations of this expression. The
variation of this transition rate ratio as a function of
temperature has been presented in Fig.3.

\begin{figure}
\includegraphics[width=8.0 cm, height = 7.2cm]{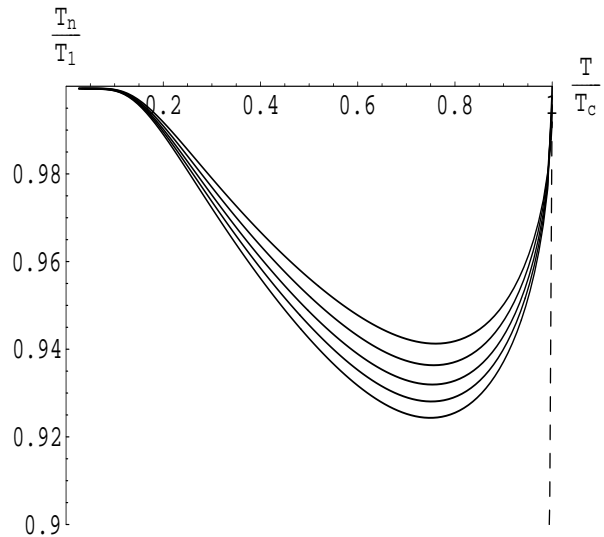}
\caption{Temperature dependence of low-frequency absorption
process obeying case I scattering coherence factor. The decreasing
order of the minima is in the decreasing order of energy gap. The
dashed curve represents corresponding calculation for BCS case.}
\end{figure}
There is a significant difference in this graph as compared to the
result based on BCS theory. At $T\simeq0$, there is a finite
contribution in the scattering rate whereas in the BCS case it is
zero. The alpha fermion does not contribute around this
temperature because there will be no fermions available to
contribute for scattering since all fermions will be frozen in the
low energy region. Mathematically it is due to the delta function
behavior of the differential of the distribution function at
energy equal to zero so that for integration all the higher energy
states will be irrelevant. So the finite contribution is from the
gapless fermions only. For the gapless fermion, at $E\simeq0$,
even though the density of states is comparable to, but slightly
greater than the density of states of normal fermions at fermi
level, the scattering rate ratio is close to but less than one.
This effect is due to the coherence factor which will be less than
one and dominates the effect of the density of states ratio.

As the temperature is increased the scattering rate is decreased.
It is definitely due to the effect of the coherence factors. The
logic is the following: With the increase in temperature, higher
energy states which are close to the gap will also be relevant and
we know there the effect of the coherence factors will be bigger.
Even if the ratio of the density of states is very high in this
energy range, the product of the density ratio and differential of
the distribution function will not be enough to overcome the
decreasing effect of the coherence factors. If the temperature is
still increased, higher energy states will also be excited. The
delta function then broadens out and the coherence factors
increase slowly up to $1$ for energy less than zero but decrease
to zero for energy greater than zero. The density of states ratio
becomes almost constant for energy less than zero but enhances for
positive energy, and the alpha fermion starts to contribute but
very weakly. For the alpha fermion the coherence factor has the
most negative effect at the diverging edge of the density of
states. The future of the scattering rate is determined by the
competition between all of these effects. The result is that the
rate ratio decreases until the temperature is close to $0.8T_{c}$
and starts to increase until it is one at $T_{c}$. From the
numerical calculation we see that the alpha fermion contribution
is responsible for the increase in the scattering rate. The curve
shows similar behavior as in BCS at around $T_{c}$ but as we
lowered the temperature it does not go to zero as in the BCS case
since in the BCS case no fermion would be available for scattering
at low temperature. Here, however, the beta fermion contribution
is comparable to normal metal scattering. So if we reduced the
temperature from $T_{c}$ we see that the initial reduction on the
scattering rate is due to the alpha fermions which are again BCS-
like and later, the increase is due to the beta fermions which are
gapless.

\section{\label{sect-theoryfour} NUCLEAR RELAXATION}

The matrix elements for nuclear-spin relaxation by interaction
with quasi-particles have the case II coherence factors, which
corresponds to the constructive interference in the relevant
low-energy scattering process. The scattering coherence factor
will have the lower sign. In the limit of $\hbar\omega\rightarrow
0$ the ratio of transitioin rate $(T_{n}/T_{1})$ can be expressed
as,

\begin{equation}
\begin{split}
\frac{T_{n}}{T_{1}}
=\int_{-\infty}^{\frac{JM}{2}-\Delta}{(1+\frac{\Delta^{2}}{({\frac{JM}{2}}-E)^{2}})}
{(\frac{N^{\beta}(E)}{N(0)})^{2}}{(\frac{-\partial{f(E)}}{\partial{E}})}dE\\
 +\int_{\frac{JM}{2}+\Delta}^{\infty}{(1+\frac{\Delta^{2}}{(E-\frac{JM}{2})^{2}})}
{(\frac{N^{\alpha}(E)}{N(0)})^{2}}{(\frac{-\partial{f(E)}}{\partial{E}})}dE.
 \label{nuclearrelaxation}
 \end{split}
\end{equation}

The variation of this transition rate ratio as a function of
temperature has been presented in fig.4.

\begin{figure}
\includegraphics[width=8.0 cm,height=6.5cm]{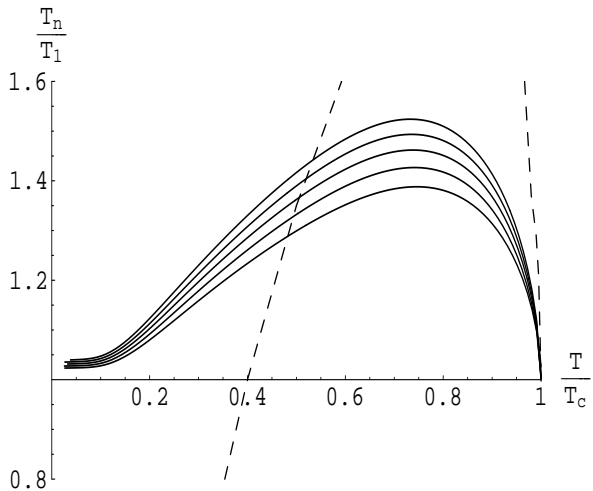}
\caption{Temperature dependence of the low-frequency absorption
process obeying case II scattering coherence factors. The
increasing peaks correspond to increase in the magnitude of the
gap. The dotted curve corresponds to the BCS case.}
\end{figure}
Reasoning similarly as above, the scattering rate is almost equal
but greater than one at around zero temperature. Only the beta
fermions contribute at low temperature since there will be no
fermions in the upper band which can contribute to scattering. The
coherence factor has maximum positive effect at
$\frac{JM}{2}-\Delta$ for the beta fermion and
$\frac{JM}{2}+\Delta$ for the alpha fermion. The initial rise in
the ratio of the scattering rate, at a temperature close to but
less than the critical temperature, is due to BCS-like alpha
fermions. By around $0.8T_{c}$ the rate is maximum and starts to
decrease, due to the freezing out of fermions in the beta fermion
band. The  finite scattering ratio at lower energy is due to the
gapless nature of the beta fermions. We solved the coupled
equations $[5,6]$ for $M$ and $\Delta$ for different values of the
interaction parameters $J$ and $g$. We refer the reader to
ref.\cite{jackiewicz} for the details. In fig.4 the scattering
rate corresponds to the pairs of $J$ and $g$ for which $M$ is
small and remains almost constant, and $\Delta$ increases faster.
And for those values of $M$ and $\Delta$, we observed that the
increase in relaxation rate ratio from around 1.35 to around 1.55
is due to the increase in the value of the gap. We see that the
peak is reduced considerably compared to the BCS case and it is
due to the presence of the finite magnetization. The increased
magnetization reduces the density of $\alpha$-fermions, the
magnitude of the coherence factors and the probability
distribution function. We will show the effect of high
magnetization explicitly in what follows. But here we want to draw
attention to the fact that we observed the coherence peak which is
a signature of an s-wave pairing mechanism.

In Fig.5, we have presented the effect of magnetization on the
nuclear relaxation rate. We again solved the coupled Eqs.$(5,6)$
for $M$ and $\Delta$ for different values of $J$ and $g$. Fig.5
corresponds to the scattering rate for pairs of $J$ and $g$ for
which $M$ is large and increases faster with the chosen values of
$J$ and $g$.
\begin{figure}
\includegraphics[width=8.0 cm,height=6.6cm]{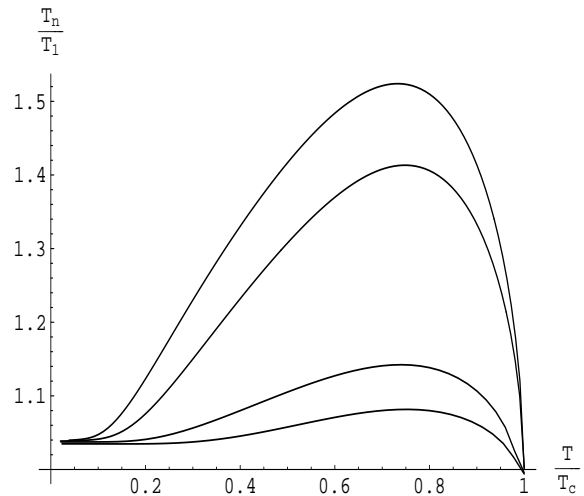}
\caption{The effect of magnetization on temperature dependence of
low-frequency absorption process obeying case II scattering
coherence factors. The reduction in peak corresponds to increase
in magnetization. The interaction parameters are $J=1.01$ fixed
and $g=1.082,1.25,1.5,2.0$ in order of reducing peak, where an
increase in g corresponds to an increase in M.}
\end{figure}

From the graph we can easily infer that even for a finite
magnetization we can get a finite peak in the relaxation rate. For
low magnetization the peak is strong and hence we can argue that
the superconductivity is dominant. More fermions are forming pairs
than aligning in certain direction to give strong magnetization.
But with the increase in magnetization the peak has been reduced.
This can be understood by arguing that when the magnetization is
increased, the fraction of fermions which are giving rise to
ferromagnetism will increase with the ultimate effect of
decreasing the fraction of fermions that form Cooper pairs. We can
increase the temperature to get more fermions to form more pairs,
but that will still hurt the paring since the thermal energy will
overcome the binding energy of the pairs. Since the coherence peak
is the inherent property of BCS pairing, once the pairing is less
effective the peak will be reduced. Nevertheless, there are always
a finite number of fermions which are forming the pairs. So we
want to put forward the idea that it is natural to have a reduced
relaxation due to the intrinsic magnetization, but there should in
s-wave superconductors always survive Cooper pairs to give an
enhanced relaxation rate, no matter how small it may be. In
addition, for the time being we believe that the mechanism of
superconductivity is s-wave and are optimistic about observing the
coherence peak in a ferromagnetic superconductor, as will be seen
in a later section.

\section{\label{sect-theoryfive} ELECTROMAGNETIC ABSORPTION}

Unlike the case of nuclear relaxation, it is now possible to
utilize large enough frequencies to allow the quasi-particles to
absorb energy, but the absorption process will now be different
compared to the case of
BCS. Now there are fermions to absorb any finite energy of
electro-magnetic wave which was already revealed
in the calculation of the nuclear relaxation and ultrasonic
attenuation where the whole curve corresponds to low energy
absorption.

The electromagnetic absorption is proportional to the real part of
the complex conductivity which we can directly compute from
Eq.(19), with the Fermi function being either zero or one at zero
temperature. The coherence factor will be the one for the pair
creation or annihilation process.  We will take the corresponding
component of coherence factors appropriate for the involved band.
For different values of incident frequency, the limit of
integration and the expression to evaluate the absorption will be
different. For two cases, namely for $\hbar\omega\leq
\frac{JM}{2}-\Delta$, and $\frac{JM}{2}-\Delta
\leq\hbar\omega\leq\frac{JM}{2}+\Delta$, we use,

\begin{equation}
\begin{split}
\frac{\sigma_{1}}{\sigma_{n}}=
\frac{1}{\hbar\omega}\int(1\mp\frac{\Delta^{2}}{(\frac{JM}{2}-E)(\frac{JM}{2}-E')})\\
 \frac{N^{\beta}(E)}{N(0)} \frac{N^{\beta}(E')}{N(0)}dE.
 \end{split}
\end{equation}
where the limit of integration for the first case is from
$-\hbar\omega$ to $0$, where as the limit of integration for the
second case is $-\hbar\omega$ to
$-\hbar\omega+\frac{JM}{2}-\Delta$. These two integrals correspond
to intra band transition in the beta fermion band. In the second
case the incident radiation will encounter the effect of the gap.
Still, since $\hbar\omega\leq\frac{JM}{2}+\Delta$, the alpha
fermion does not contribute to absorption. For
$\hbar\omega\geq\frac{JM}{2}+\Delta$,

\begin{equation}
\begin{split}
 \frac{{\sigma_{1}}^{\beta\beta}}{\sigma_{n}}=
\frac{1}{\hbar\omega}\int(1\mp\frac{\Delta^{2}}
{(\frac{JM}{2}-E)(\frac{JM}{2}-E^{\prime})})\\
{\frac{N^{\beta}(E)}{N(0)}}{\frac{N^{\beta}(E^{\prime})}{N(0)}}dE,
\end{split}
\end{equation}

\begin{equation}
\begin{split}
\frac{{\sigma_{1}}^{\beta\alpha}}{\sigma_{n}}=
\frac{1}{\hbar\omega}\int(1\pm\frac{\Delta^{2}}
{(\frac{JM}{2}-E)(E^{\prime}-\frac{JM}{2})})\\
\frac{N^{\beta}(E)}{N(0)}\frac{N^{\alpha}(E^{\prime})}{N(0)}dE,
\end{split}
\end{equation}

\begin{equation}
\frac{\sigma_{1}}{\sigma_{n}}=
\frac{{\sigma_{1}}^{\beta\beta}}{\sigma_{n}} +
\frac{{\sigma_{1}}^{\beta\alpha}}{\sigma_{n}},
\end{equation}
where $\frac{\sigma_{1}}{\sigma_{n}}$ is the total conductivity
and is the sum of conductivity due to transition from beta to beta
$\frac{{\sigma_{1}}^{\beta\beta}}{\sigma_{n}}$ band and from beta
to alpha band $\frac{{\sigma_{1}}^{\beta\alpha}}{\sigma_{n}}$. The
limit of integration for the first integral is from $-\hbar\omega$
to $-\hbar\omega+\frac{JM}{2}-\Delta$ and for the second integral
is $-\hbar\omega+\frac{JM}{2}+\Delta$ to $0$. The upper sign is
for case I and the lower sign is for case II. The second integral
is the contribution of the alpha fermion which reduces to a BCS
like expression if M goes to zero. The first term still gives
intra-band transitions and the second term gives inter band
transitions. It is very easy to see that the true gap is only
$2\Delta$. We evaluated the ratio of conductivity for both the
cases. The variation of this transition rate ratio as a function
of temperature at fixed value of $J, g$ has been presented in
Fig.(6)and Fig.(7) for case I and case II respectively.

 \begin {figure}
\includegraphics[width=9.50cm,height=6.5cm]{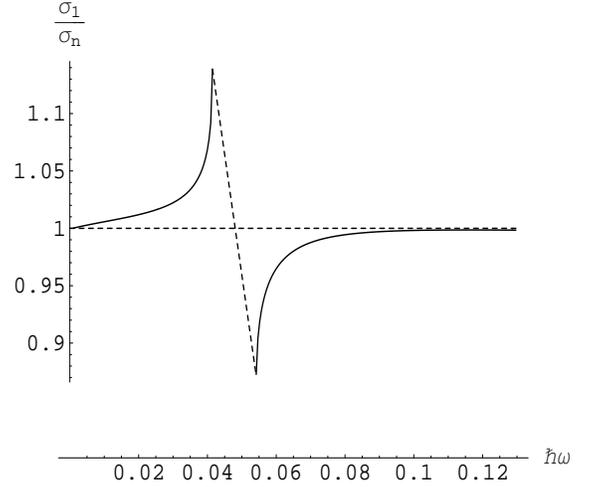}
\caption{Frequency dependence of absorption process obeying case I
coherence factor.}
\end{figure}

\begin {figure}
\includegraphics[width=9.2cm,height=6.5cm]{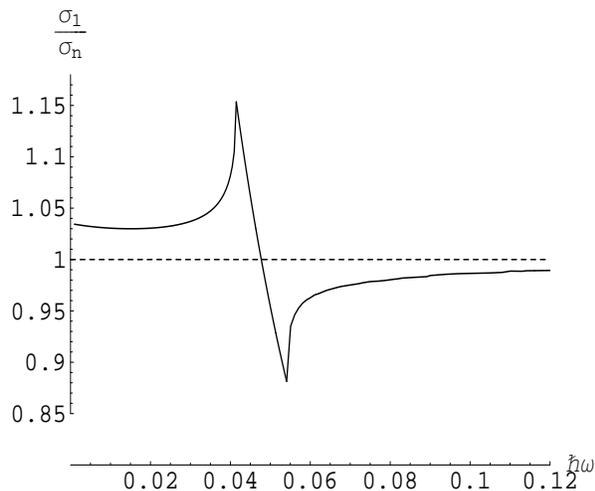}
\caption{Frequency dependence of absorption process obeying case
II coherence factor.}
\end{figure}
Both the cases show almost similar behavior except for some
peculiarity in each case. For both the cases, the absorption rate
is never zero because of the presence of gapless fermions. It
increases proportionally to the increase in density of states of
the beta fermion if the incident energy is enough to excite
fermions to a higher energy region of the beta fermion band. The
case I has lower absorption rate than case II if the radiation
energy is less than the maximum possible value of the beta fermion
energy because of the coherence factors. Once the incident energy
is higher than the maximum possible value of the beta fermion
energy and less than minimum value of the alpha fermion energy,
the effect of the gap will come into play and the absorption rate
drops sharply. But this will not go to zero since still there is
the presence of intra band transitions in the beta fermion bands.
For energy higher than the minimum value of the alpha fermion
energy the absorption rate will start to increase again and reach
the value of the normal fermion absorption rate in the high energy
limit. To be precise,in this energy limit the ratio of the rate
for case I will be higher and saturate out faster than that of
case II. If we compare with the BCS case, the behavior is similar.
The only difference in the BCS case is that for case I the ratio
has a discontinuous jump when the energy of the electromagnetic
radiation is bigger than the gap energy and has the absorption
rate much bigger than that of normal fermions. For the case II the
ratio starts to increase from zero with a faster rate in the BCS
case, but here it starts at a finite value and rises more slowly.
Here, both the inter band and intra band transitions will
contribute for absorption. Another important point, unlike the BCS
case, there will be no delta function-like behavior of the
conductivity corresponding to zero frequency (DC conductivity). In
the BCS case the lower band will be at an energy which is less
than zero, hence occupied, and has infinite density of states
which contribute an infinite conductivity but here at zero
temperature, the system will never have infinite density since an
enhanced density corresponds to positive energy and will be
unoccupied at zero temperature. Interestingly enough, both the
cases satisfy the optical sum rule in the more usual manner even
though there is a very slight discrepancy in the area bounded by
the curve, that we guess might be due to the presence of a
diverging term (density of states) at one extreme limit of
integration.

\section{\label{sect-prediction} PREDICTION AND DISCUSSION FOR $UGe_{2}$}

Two of the present authors \cite{jackiewicz} have studied some
physical properties of $UGe_{2}$ using the same model
\cite{karchev2001} as we have used here. In the study of the
specific heat of this system, they found that their calculation
fits well the data observed experimentally \cite{Tateiwa2001}. The
interaction parameters used in that calculation are $J=1.001$ and
$g=1.1$. We did the calculation of the relaxation rate ratio with
respect to the normal state contribution with the same interaction
parameter. The graph has been presented in Fig.8. It is clear from
the figure that the enhancement in the relaxation rate ratio is
very small, only around $7 \%$ over the normal metal value. We
think it might be hard to see this small enhancement
experimentally.

\begin {figure}
\includegraphics[width=8.0cm,height=6.5cm]{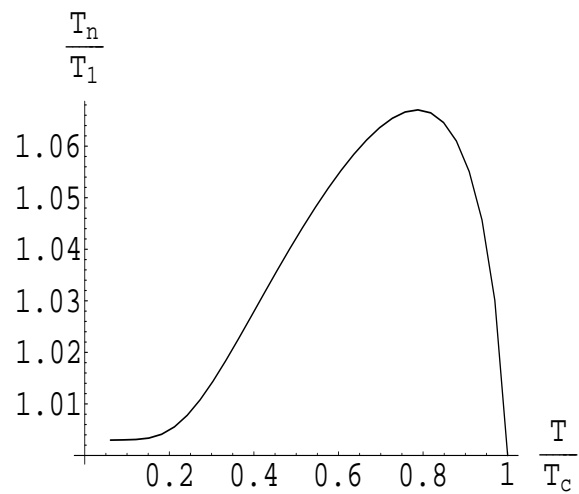}
\caption{The nuclear relaxation rate for $J=1.001$ and $g=1.1$.
The coherence peak has a $7 \%$ enhancement over the normal metal
value.}
\end{figure}
To give a quantitative feeling about the predicted values and the
experimental values of some properties we have presented a table
which shows the maximum percentage variation of these physical
parameters with respect to the normal metal values for $J=1.001$
and $g=1.1$.
\begin{center}
  \begin{tabular}{|c|c|c|c|c|}
    \hline
     \bf{property} & \bf{predicted} & \bf{experimental} \\
    \hline
    specific heat & $(\Delta{C_{V}}/C_{V})=20 \%$ & $(\Delta{C_{V}}/C_{V})=20 \%$ \\
    \hline
    nuclear relax. & $T_{n}/T_{1}=7 \%$ & $T_{n}/T_{1}\leq15\%$ \\
    \hline
    ultras. atten. & $T_{n}/T_{1}=1.5 \%$  & $T_{n}/T_{1}=?$ \\
    \hline
    electromag. abs. & $\sigma_{1}/\sigma_{n}=4 \%$ & $\sigma_{1}/\sigma_{n}=?$ \\
    \hline
  \end{tabular}
\end{center}
It is clear from the data presented in the table that the specific
heat jump is greatly reduced compared to the BCS case with a
maximum enhancement of only $20 \%$ over the normal value. We
assume that the maximum enhancement in the coherence peak height
is less than $15 \%$, explained below, which is very small
compared to the more than $100 \%$ rise over the normal value in
the BCS case. The theoretical study of other properties like
ultrasonic attenuation and electromagnetic absorption also show
very feeble variation from the normal state value.

Experimental data for $1/T_{1}$ as a function of $T$ at pressures
around $13Kbar$ is available in Ref.\cite{kotegawa2003}. To
compare our prediction with those data, we evaluate our expression
and present the result in Fig.9.

\begin {figure}
\includegraphics[width=8.0cm,height=6.5cm]{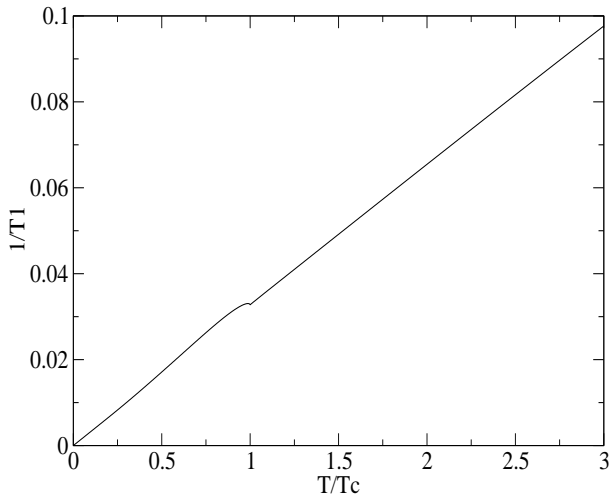}
\caption{The nuclear relaxation rate for $J=1.001$ and $g=1.1$.
The coherence peak is very small.}
\end{figure}

We can see that the coherence peak is very small using the
interaction parameters that have been used for all of the
calculations, including the specific heat. In the data of Ref.20,
the peak is unseen. However, the plot there is on a log-log scale,
which could reduce its appearance based on a reproduction of that
data that we have done. The point is that the evidence is not
clear that there is not a coherence peak in the experiments that
have been carried out so far, and there needs to be a more
detailed exploration with smaller temperature steps around the
superconducting transition. We believe there is a very small, yet
noticeable upturn in the relaxation rate.

Furthermore, in the experiments by the same group on the s-wave
superconductor $MgB_{2}$ for $1/T_{1}$ \cite{kotegawa2001}, the
plot looks incredibly similar to the one for $UGe_{2}$. The peak
is slightly more noticeable for this well-studied material, but
there is no background ferromagnetism present to suppress it as
much as in the case of $UGe_{2}$.

Another interesting point of our model with respect to experiment
can be seen in a paper soon to be published \cite{kotegawa2004}
where $1/T_{1}T$ for $UGe_{2}$ is studied well below the
superconducting transition temperature. It is seen that this value
holds very constant, independent of pressure and temperature. One
possible reason surmised for this behavior is the likely presence
of low lying gapless excitations, which is an inherent property of
the coexistent phase of our model.

\section{\label{sect-conclu} CONCLUSION}

We closely studied the density of states, ultrasonic attenuation,
nuclear relaxation and electromagnetic absorption for a
ferromagnetic superconductor using a mean field theoretical
approach where superconductivity is due to s-wave pairing and the
magnetism is due to spontaneously broken spin rotation symmetry.
Due to the finite density of states of gapless fermions, the low
temperature behavior of the properties we studied is dominated by
normal metal like behavior. At temperature less than the
superconducting transition temperature, our study supports the
presence of superconductivity of an s-wave nature. Namely, there
is the presence of a Hebel-Slichter peak. The peak is reduced due
to the presence of a ferromagnetic background. There is a slight
decrease in the ultrasonic attenuation rate ratio when the
temperature is less than the transition temperature. The
electromagnetic absorption shows not much difference in the
conductivity for either case. Nonetheless, it shows some
interesting behavior compared to the result in the BCS case. The
optical sum rule is followed by both cases in a more general way.
We are partially successful in describing the experimental
observation of the temperature dependence of the relaxation rate
in supercoducting $UGe_{2}$.

We graciously thank to Kotegawa group for providing us their
experimental data for $UGe_{2}$. This work was done with the
support of DOE/DEFG0297ER45636.

\bibliographystyle{apsrev}
\bibliography{mypaper}
\end{document}